\documentclass[aps,pra,twocolumn,showkeys,nofootinbib,superscriptaddress,dvipsnames]{revtex4}
\pdfoutput=1

\usepackage[utf8]{inputenc}
\usepackage{graphicx}
\usepackage{graphics}
\usepackage{epsfig}
\usepackage{lscape}
\usepackage{color}
\usepackage{multirow}
\usepackage{hyperref}
\usepackage{breqn}
\usepackage{physics}
\usepackage{subfiles}
\usepackage{ragged2e}
\usepackage{amssymb}
\usepackage{soul}
\usepackage[dvipsnames]{xcolor}

\newcommand{\Chemie}{
Department Chemie,
Johannes Gutenberg-Universit\"at, Fritz-Strassmann Weg 2, 55128 Mainz, Germany \\
}

\newcommand{\HIM}{
Helmholtz-Institut Mainz,
Staudingerweg 18, 55128 Mainz, Germany \\
}

\newcommand{\Physik}{Institut für Physik, Johannes Gutenberg-Universit\"at, Staudingerweg 7, 55128 Mainz, Germany\\}

\newcommand{\GSI}{
GSI Helmholtzzentrum f\"ur Schwerionenforschung,
Planckstrasse 1, 64291 Darmstadt, Germany\\
}

\begin{document}
\title{Laser Resonance Chromatography: First Commissioning Results and Future Prospects \footnote[3]{This publication uses data collected within the framework of the PhD thesis of EunKang Kim.}}

\author{EunKang Kim}
\affiliation{\Chemie} \affiliation{\HIM}

\author{Biswajit Jana}
\affiliation{\Chemie} \affiliation{\HIM}

\author{Aayush Arya}
\affiliation{\Physik} \affiliation{\HIM}

\author{Michael Block}
\affiliation{\Chemie} \affiliation{\HIM} \affiliation{\GSI} 

\author{Sebastian Raeder}
\affiliation{\GSI} \affiliation{\HIM}

\author{Harry Ramanantoanina}
\affiliation{\Chemie} \affiliation{\HIM}

\author{Elisabeth Rickert}
\affiliation{\Chemie} \affiliation{\HIM} \affiliation{\GSI}

\author{Elisa Romero Romero}
\affiliation{\Chemie} \affiliation{\HIM}

\author{Mustapha Laatiaoui\footnote[1]{Corresponding author: Laatiaoui@uni-mainz.de}} 
\affiliation{\Chemie} \affiliation{\HIM} \affiliation{\GSI}

\date{\today}

\begin{abstract}
We report first results obtained during the commissioning of the Laser Resonance Chromatography (LRC) apparatus, which is conceived to enable atomic structure investigations in the region of the heaviest elements beyond nobelium. 
In our studies we first established optimum conditions for the operation of the different components of the setup, including the radio-frequency quadrupole ion buncher and the cryogenic drift tube, which was operated with helium buffer gas at relatively low electric fields. We used laser ablated hafnium, lutetium, and ytterbium cations to assess the chromatography performance of the drift tube at a gas temperature of $295$K. Arrival time distributions of singly charged lutetium revealed two distinct ion mobilities of this ion in the ground and metastable state in helium with a relative difference of about $19$\%. By using $^{219}$Rn ions from a $^{223}$Ra recoil source the overall efficiency of the apparatus is found to be $(0.6\pm0.1)$\%. The findings help to establish LRC on lutetium, which is the lighter chemical homolog of lawrencium.  
\end{abstract}

\maketitle

\section{Introduction}
Since the discovery of the first transuranium elements in the 1940s, efforts have been made to create ever new heavier elements \textcolor{black}{\cite{Muenzenberg:2015,Oganessian:2015}} and to study their basic properties \cite{Herzberg:2008,Block:2021} such as decay modes and half-lives. The main aim is to know under which conditions these synthetic atoms can be produced and how they behave chemically \cite{Tuerler:2015,Duellmann:2017}. 
The atomic structure can be modeled with sufficient accuracy using modern \textit{ab initio} calculations \cite{Eliav:2015,Dzuba:2017} and experimentally scrutinized by means of optical spectroscopy \cite{Backe:2015}. The measured spectral lines, together with the theoretical calculations, thus provide a solid basis for studying the influence of relativistic and many body effects on the atomic structure of these heavy elements. 
\textcolor{black}{Presently, we know the existence of 118 chemical elements}. However, the heaviest element for which optical spectroscopy has been possible to date is nobelium (No), with atomic number $Z=102$ \cite{Laatiaoui:2016}. For the superheavy elements (SHEs) beyond nobelium, direct experimental investigation of the atomic structure is a challenge, primarily due to the extremely low production yield and the very short half-life of these elements. Conventional spectroscopy techniques based on fluorescence detection are no longer suitable because they do not have the sensitivity required to study SHEs. For the studies carried out on nobelium \cite{Laatiaoui:2016}, resonance ionization spectroscopy has proven to be sufficiently sensitive. Whether it can also be used for the SHEs remains an open question, despite the continuous efforts to investigate the element lawrencium (Lr, $Z=103$), the heaviest actinide element~\cite{Warbinek:2022}.

The recently proposed Laser Resonance Chromatography (LRC) \cite{Laatiaoui:2020a,Laatiaoui:2020b,Ramanantoanina:2021,Visentin:2024} could remedy this by providing sufficient sensitivity for the study of superheavy ions and overcoming the difficulties associated with other methods. 
For lawrencium, for example, the procedure is as follows: Starting from Lr$^+$ ions in their atomic ground state ($7s^2$ $^1$S$_0$), one tries to populate predicted electronic states \cite{Kahl:2019,Ramanantoanina:2022} by means of laser resonant excitations. If the excitation process is successful, a low-lying metastable state ($6d 7s$ $^3$D$_1$) is subsequently populated very efficiently, and one obtains a statistical ensemble of Lr$^+$ ions in different electronic states that can be unambiguously detected by electronic state chromatography \cite{Kemper:1990,Ramanantoanina:2023}. 
To this end, the ions are sent under the influence of an external electric field $E$ through a drift tube filled with helium (He) gas in order to record their arrival time distribution (ATD). The state separation occurs within the drift tube since the drift velocity $v_{\text{d}}$ of the ions is proportional to the ion mobility $K$ according to \cite{Viehland:2018} 
\begin{equation}
    v_{\text{d}}= K\cdot E, 
    \label{Eq:velocity}
\end{equation}
which in turn depends on the electronic configuration of the corresponding states \cite{Laatiaoui:2012,Ramanantoanina:2023}. If the frequency of the laser light does not match the resonance frequency of the ground state transition, only a single peak in the ATD can be expected corresponding to the ground state occupation. 
\textcolor{black}{Conversely, when the laser is in resonance with the ground state transition, the picture} changes abruptly and a new additional peak appears in the arrival time distribution, which makes up the LRC signal~\cite{Laatiaoui:2020b}. 

\begin{figure*}[htb!]
\centering
\includegraphics[width=\textwidth]{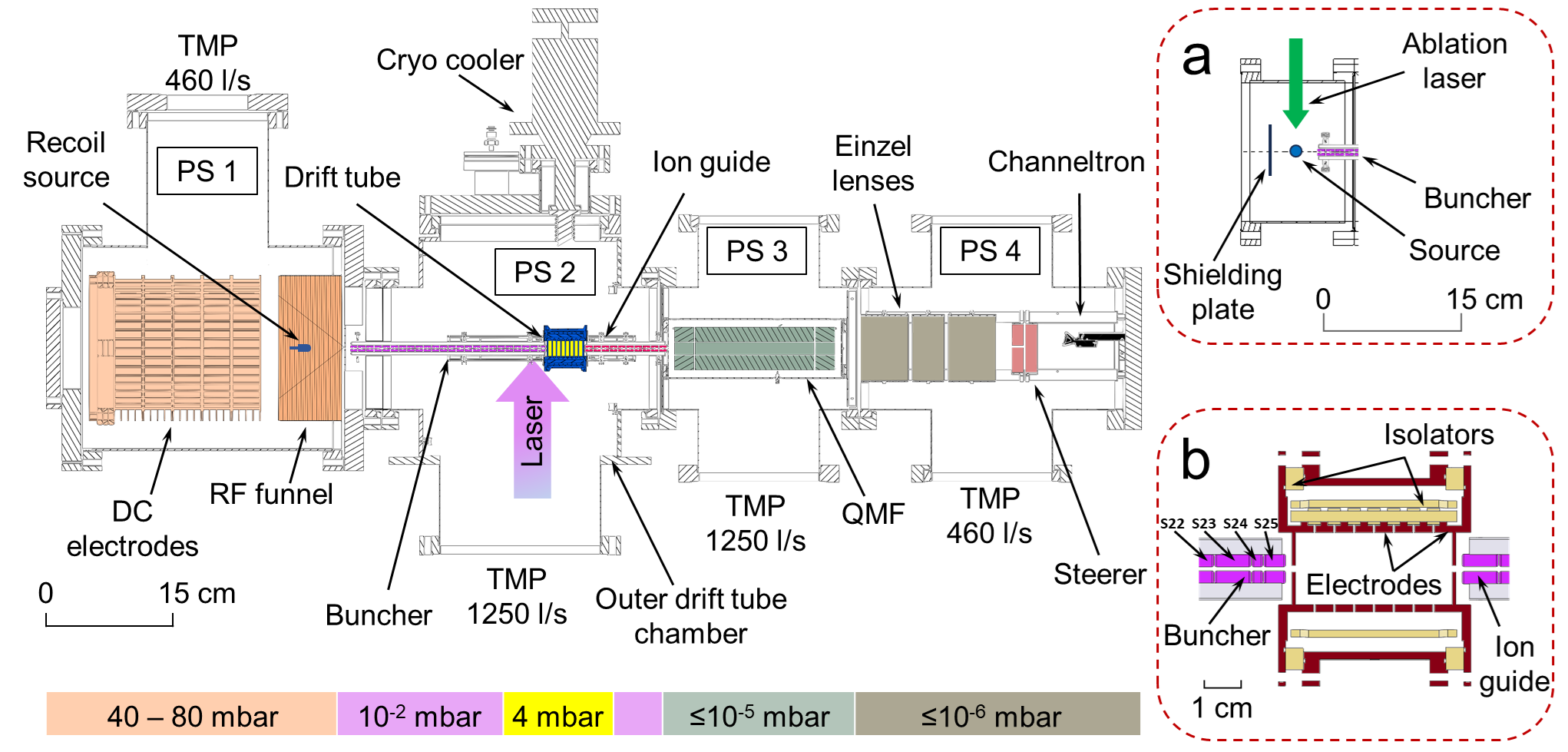}
\caption{Schematic overview of the LRC apparatus. For efficiency measurements, a $^{223}$Ra recoil source was inserted into the stopping cell close to the exit nozzle. Typical pressure ranges are indicated for each pumping section of the apparatus. \textbf{a}: Cross piece, which replaced the stopping cell to enable the operation of the laser ablation source in vacuum. \textbf{b}: Drift tube connected to the RF buncher (left) and to the ion guide (right). See text for more information.}
\label{fig:LRC_setup}
\end{figure*}  

In a recent work \cite{RomeroRomero:2022}, the LRC setup was introduced and results of ion trajectory simulations were presented for the drift of scandium ions in helium buffer gas, \textcolor{black}{under different ratios of electric field to gas number density ($n_0$)} or the so-called reduced fields $E/n_0$. 
In this paper we provide an overview of the LRC apparatus and its working principle with a focus on the radio-frequency (RF) structures used to control and trap the ion ensembles. 
We report results from commissioning experiments, which focused on optimizing the time resolution of the chromatography stage 
\textcolor{black}{utilizing three metal cations of completely different ground-state configurations, Yb$^+$ ($4f^{14} 6s$), Lu$^+$ ($6s^2$), and Hf$^+$ ($5d 6s^2$).  
Finally, we report the overall efficiency of the device and 
conclude with an outlook on the laser resonance chromatography of Lu$^+$ as an intermediate step towards LRC on its chemical homolog, Lr$^+$.} 

\section{Experimental apparatus}
The experimental apparatus is schematically shown in Fig. \ref{fig:LRC_setup}. It consists of four main vacuum sections (labelled PS$1-4$).
As SHEs are produced in nuclear fusion reactions at high-energy beam facilities, the fast reaction products \textcolor{black}{must be thermalized in a gas-stopping cell.}
Such a buffer gas stopping cell makes up the first vacuum section (PS1) of the LRC apparatus. Currently, we are using the former room-temperature stopping cell of the SHIPTRAP project \cite{Neumayr:2006a} for this purpose. 
The stopping cell is operated at pressures between $40$ to $80\,$mbar of He gas having a purity of at least $99.996\,$\%, that we further purify utilizing a gas getter (type: MonoTorr, PS4-MT3-R-2). The cell consists of a cage with five ring electrodes of $160\,$mm inner diameter, which together extend over a total length of $180\,$mm, encapsulating a relatively large stopping volume. The cage electrodes directly lead to the RF funnel. The latter consists of $40$ one-mm thin electrodes of converging inner diameters, and serves to focus the accumulated product ions into an ion extraction nozzle of $0.6\,$mm throat diameter. 
\textcolor{black}{For experiments that rely on offline sources} of ions, we installed a flange (DIN 100 CF) at the entrance window of the cell, on which a radioactive recoil source was mounted.
The reader is referred to Refs. \cite{Neumayr:2006a,Neumayr:2006b} for further technical details of this stopping cell.
The DC voltages of the cage electrodes in the cell are set individually. The funnel is operated with RF voltages above $100\,$V$_{\text{pp}}$ at a frequency of $980\,$kHz. The funnel electrodes are connected by a resistor chain, such that an ion guiding electric field can be easily generated along the funnel structure by setting suitable DC voltages for the first and the last funnel electrode.

The second section (PS2) incorporates an RF ion buncher, a cryogenic drift tube, and an RF ion guide, all of which together form practically the heart of the LRC apparatus, \textit{cf.} Fig.~\ref{fig:LRC_setup}(b). The RF structures operate based on the principle of a linear Paul trap and serve to bunch the ions in measurement cycles and guide them through the second section. This section is evacuated by a turbo-molecular pump (TMP, type: Edwards, STPA-1603C) of a maximum pumping speed of $1250\,$l/s for He gas. 

The third vacuum section (PS3) is a four-way standard cross (DIN 150 CF) that has been shortened on one axis to a length of about $226\,$mm to accommodate a quadrupole mass filter (QMF) assembly (type: Extrel). This section is evacuated by another TMP (type: Edwards, STPA-1603C, $1250\,$l/s He pumping speed) to pressures below $10^{-5}\,$mbar during QMF operation. The QMF assembly consists of four main QMF rods with a diameter of $19\,$mm and a length of $200\,$mm including pre- and post-filters, so called Brubaker lenses \cite{Brubaker:1968}. It is used together with a controller (type: Extrel, QPS 500) to select mass-to-charge-ratios ($M/Z$) of interest for transmission of singly charged ions in the mass range $1-500\,$ atomic mass units (amu). The Brubaker lenses act thereby as high-pass filters for the ion's $M/Z$ and help to increase the ion transmission through the QMF structure. A stainless steel orifice with a $3\,$mm diameter opening separates the second and third pressure sections to allow efficient differential pumping. 
The QMF assembly is operated at a radio-frequency of $1.2\,$MHz and can be set on a floating voltage (called \textit{axial voltage} hereafter), for better tuning of ion velocities and ion transmission through the filter. We use analog interfaces to control the axial voltage, the mass-to-charge ratio, and the mass resolving power of the QMF. During the measurements presented here, the axial voltage was always kept at $0$ V, while the mass resolving power was adjusted upon the needs for each mass-to-charge range investigated here.

The fourth and last section (PS4) is evacuated separately by a TMP (type: Edwards, STP-451) such that a background pressure in this section of $\leq10^{-6}\,$mbar can be maintained. The section contains an einzel lens constellation \cite{Liebl:2008} consisting of three aluminum ring electrodes with the same inner diameter of $80\,$mm but different lengths, namely $60\,$mm, $40\,$mm, and $60\,$mm in a row, separated by $6\,$mm gaps. The lens constellation is followed in series by two pairs of aluminum half-ring plates (steerer 1 \& 2) with a diameter of $60\,$mm and a length of $15\,$mm to direct the transmitted and focused ion beam into a channel electron multiplier detector (channeltron, type: Dr. Sjuts, KBL 15RS). 
In all our experiments, a dry pump (type: Edwards, iXL600, $600\,$m$^3$/h pumping speed) is used as a backing pump. 

\textcolor{black}{For simplicity, we no longer use an intermediate vacuum section with an extraction RFQ as envisaged in Ref.~\cite{Romero:2022} because the RF buncher provides extraction, gas cooling, and bunching of ions all together, while overcoming the need for additional gas injection into the buncher section for optimal performance. The current arrangement} of pumping sections as shown in Fig.~\ref{fig:LRC_setup} makes up the setup for online experiments and is used in the present study to determine the ion transmission efficiency of the apparatus utilizing a $^{223}$Ra recoil source. 
However, most of the results reported here were obtained without utilizing the stopping cell. Instead, a special cross piece (DIN 150 CF) was connected to the second vacuum section in order to use a laser ablation source for ion production, \textit{cf.} Fig.~\ref{fig:LRC_setup}(a). 
Hereafter, we refer to the studies using the two types of sources, the laser ablation source with metal foil samples to generate Yb$^+$, Lu$^+$, and Hf$^+$ ions as \textit{ablation studies} and the $^{223}$Ra recoil source to characterize the efficiency of the system as \textit{efficiency studies}. 
A more detailed description of the most important experimental components and configurations is given below. 

\subsection{Voltage configurations} 
\textcolor{black}{With the exception of the channeltron bias voltages}, all DC voltages for the different electrodes are provided from a universal multichannel power supply (CAEN, type: SY 5527LC) in conjunction with four 12-channel high voltage boards for positive voltages (type: AG 538DP, $+1.5\,$kV, $10\,$mA) and one 12-channel board for negative voltages (type: AG 538DN, $-1.5\,$kV, $10\,$mA).
To operate the buncher and the ion guide at the required RF voltages, we developed electric circuits equipped with ferrite-core-based coils.
In general, both the buncher as well as the ion guide are operated at RF voltages $\leq30\,$V$_{\text{pp}}$ such that two arbitrary function generators (type: Aim TTi TGF4042) are sufficient to drive both required resonance radio frequencies, respectively $820\,$kHz and $980\,$kHz, without utilizing dedicated RF amplifiers. The guiding electric field is defined by the voltage gradients, and must be tuned for the needs of each ion species. Table~\ref{tab:Voltages1} shows the optimum voltage configurations for highest count rates depending on which ion source is used, \textit{cf.} Fig.~\ref{fig:LRC_setup}.

\begin{table}[htb!]
    \caption{Standard operational parameters for ablation and efficiency studies utilizing the three-metal ablation source and the $^{223}$Ra recoil source, respectively.}
    \centering
    \begin{tabular}{l|cc}    
         \hline \hline
         Electrode                           & Ablation                & Efficiency \\
                                             & studies                 & studies \\
         \hline
         \multicolumn{3}{c}{\textbf{DC voltages:}} \\
         Recoil source                       & ---                     & $+196\,$V \\
         Cage electrode 1                    & ---                     & $+196\,$V \\
         Cage electrode 2                    & ---                     & $+186\,$V \\
         Cage electrode 3                    & ---                     & $+176\,$V \\
         Cage electrode 4                    & ---                     & $+166\,$V \\
         Cage electrode 5                    & ---                     & $+156\,$V \\ 
         Funnel entrance electrode           & ---                     & $+140\,$V \\
         Funnel exit electrode               & ---                     & $+128\,$V \\
         Nozzle                              & ---                     & $+126\,$V \\
         Shielding plate                     & $+143\,$V               & --- \\
         Ablation source                     & $+143\,$V               & --- \\
         Buncher segment S1                  & $+143\,$V               & $+104\,$V \\
         Buncher segment S22                 & $+142\,$V               & $+103\,$V \\
         Buncher kicker segment S23 (high)   & $+243\,$V               & $+126\,$V \\
         Buncher kicker segment S23 (low)    & $+141\,$V               & $+101\,$V \\
         Buncher trap segment S24            & $+140\,$V               & $+100\,$V \\
         Buncher repeller segment S25 (high) & $+153\,$V               & $+102\,$V \\
         Buncher repeller segment S25 (low)  & $+130\,$V               & $ +88\,$V \\
         Drift tube entrance electrode       & $ +85\,$V               & $ +68\,$V \\
         Drift tube exit electrode           & $ +15\,$V               & $ +25\,$V\\
         Ion guide entrance segment          & $ +10\,$V               & $ +15\,$V\\
         Ion guide exit segment              & \multicolumn{2}{c}{$+5\,$V} \\
         QMF Diaphragm                       & \multicolumn{2}{c}{$0\,$V} \\
         Brubacker lens 1                    & \multicolumn{2}{c}{$0\,$V} \\
         QMF axial voltage                   & \multicolumn{2}{c}{$0\,$V} \\ 
         Brubacker lens 2                    & \multicolumn{2}{c}{$0\,$V} \\
         Einzel lens electrode 1             & \multicolumn{2}{c}{$-999\,$V} \\
         Einzel lens electrode 2             & \multicolumn{2}{c}{$0\,$V} \\
         Einzel lens electrode 3             & \multicolumn{2}{c}{$-999\,$V} \\
         Steerer 1 electrode 1 (horizontal)  & \multicolumn{2}{c}{$0\,$V} \\
         Steerer 1 electrode 2 (horizontal)  & \multicolumn{2}{c}{$0\,$V} \\
         Steerer 2 electrode 1 (vertical)    & \multicolumn{2}{c}{$0\,$V} \\
         Steerer 2 electrode 2 (vertical)    & \multicolumn{2}{c}{$-150\,$V} \\
         Channeltron entrance                & \multicolumn{2}{c}{$-3000\,$V} \\
         Channeltron exit                    & \multicolumn{2}{c}{$0\,$V} \\
         Channeltron collector plate         & \multicolumn{2}{c}{$+200\,$V} \\
         \hline
         \multicolumn{3}{c}{\textbf{RF voltages:}} \\
         Funnel                              & ---                     & $182\,$V$_{\mathrm{pp}}$ \\
         Buncher                             & \multicolumn{2}{c}{$19\,$V$_{\mathrm{pp}}$} \\
         Ion guide                           & \multicolumn{2}{c}{$16\,$V$_{\mathrm{pp}}$} \\        
         \hline \hline
    \end{tabular}
    \label{tab:Voltages1}
\end{table}

\subsection{Ion sources}
First, to determine the transmission efficiency of the apparatus, a $^{223}$Ra recoil source is used. As the recoil source emits ions at a specific decay rate, the number of ions entering the setup can be determined at any time during the experiments, provided that the initial activity is known. The $^{223}$Ra source primarily produces $^{219}$Rn recoil ions with a half-life of $T_{1/2}(^{219}\mathrm{Rn})= 3.96\,$s, which is much longer than the expected ion extraction time of several milliseconds~\cite{Neumayr:2006c} and thus best suited for transmission efficiency measurements. 
The source is first prepared via a breeding process, where $^{223}$Ra ions are gently deposited on a stainless steel hemisphere of $10$ mm diameter. The breeding process is carried out in the radiochemistry lab of the University of Mainz and takes several days, depending on how much activity the source is supposed to deliver.
Then the source is installed inside of the stopping cell by fixing it on a special electrically insulated mount on the entrance window flange, which enables
to set the source on a certain positive DC voltage for repelling ejected recoil ions. In our experiments we installed the source along the axis of the stopping cell, deep inside the conical RF funnel, at a distance of $30\,$mm before the nozzle throat.

Figure~\ref{fig:LRC_setup}(a) shows the configuration for the buncher section used during the ablation studies. The ablation source is a cylindrical plate
on top of which $0.1\,$mm thin foils ($99.9$\% purity) of Hf, Lu, and Yb are fixed. Its bias voltage is given in Tab.~\ref{tab:Voltages1}.
For ablation, we use an Nd:YAG laser (type: Continuum, Minilite II) operating at a wavelength of $532\,$nm. 
The laser is externally triggered by the experiment control and data acquisition system (DAQ, type: NI PCIe-7857) \textcolor{black}{and fires with a delay of $145\,$µs to the trigger due to its Q-switched operation.} 
Depending on experimental requirements, both the laser power and the repetition rate can be set, with the latter adjusted between $1$ and $10\,$Hz~\cite{RomeroRomero:2022}.
The laser beam can be steered utilizing a piezo-driven mirror such that different positions on the cylindrical plate can be chosen as laser beam focus. 
The ablation source can additionally be rotated around its own axis such that different foils can be targeted without changing the laser beam focus on the ablation source.

\subsection{Drift tube}
The drift tube is where the state separation takes place. 
After trapping the sample ions in a potential well, as described in the next section, different electronic states can be excited resonantly via a tunable dye laser. One laser should be sufficient in this case. More details on such a dye laser system can be found in Ref.~\cite{RomeroRomero:2022}. 
Ions in different electronic states have different collisional cross sections with helium. Therefore ions in different states drifting through the tube will undergo different number of collisions, resulting in different drift velocities. This causes them to reach the detector at state-specific time imprints and thus to have different arrival time distributions. 
It should be noted here that the isotope mass only plays a minor role and that similar arrival times can be expected for the same species in the same state.

Figure~\ref{fig:LRC_setup}(b) shows a schematic view of this drift tube. 
Due to its cooling capacities, it can also be used when cryogenic gas temperatures are required for state separation. The tube has a length of $53.5\,$mm, providing a drift length of about $45\,$mm. The drift tube consists of eight stainless steel electrodes, including the end caps, which isolate the drift tube volume from the PS2 vacuum section.
All electrodes are electrically isolated from the drift tube chamber, but are interconnected by a series of seven 1-M$\Omega$ resistors, resulting in a resistance of $7.15\,$M$\Omega$ between the end caps.
These latter caps exhibit entrance and exit orifices of $2\,$mm diameter and simultaneously serve as fixation platforms for the RF buncher from the entrance side, and for the ion guide at the exit. The drift tube chamber is copper plated from the outside and connected to a free piston Stirling cryocooler (type: CryoTel-CT, cooling capacity: $11\,$W at $77\,$K) via four copper strands with a cross sectional area of $16\,$mm$^2$. 
Moreover, the drift tube features inlet and outlet tubes for gas injection and pressure monitoring as well as high-load resistor heaters and temperature sensors for conditioning purposes. 
More technical details on the tube, including its expected performance for scandium cations can be found in Ref.~\cite{RomeroRomero:2022}.

\subsection{Miniature RF buncher}
The miniature RF buncher is a segmented quadrupole ion guide, which enables spatial ion confinement for laser spectroscopy and precise ion bunching for drift time measurements. It has a length of 243 mm and is composed of $25$ electrode segments. Each segment is composed of four rods of $3.5\,$mm diameter making a quadrupole geometry such that the distance between the opposite rods is $3\,$mm. The spacing between the segments ranges from $0.2$ to $0.5\,$mm. Figure~\ref{fig:LRC_setup}(b) shows the last four segments of the buncher close to the entrance end cap of the drift tube. Each segment is $10\,$mm long. However, the last (S25) and the second-last (S24) buncher segments are $8$ and $3\,$mm long, respectively. 
\begin{figure*}
\centering
\includegraphics[width=0.75\textwidth]{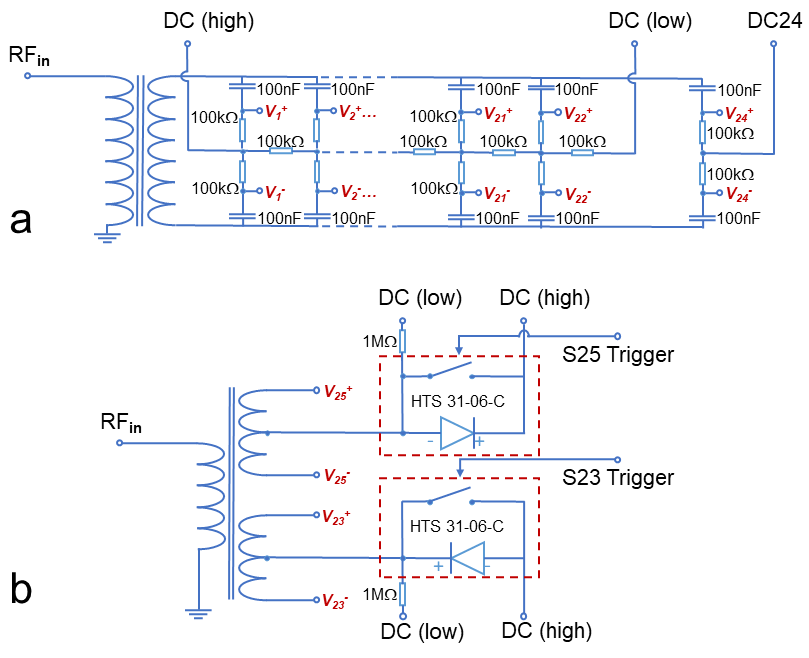}
\caption{Schematic overview of the buncher circuit for LRC operation. \textbf{a}: RF voltages of opposite phases are applied to segments S1-S22 and to the trap segment S24 through a main RF coil and $100\,$nF coupling capacitors. Different DC voltages are additionally applied to the segments via $100\,$k$\Omega$ coupling resistors to give the segment specific RF voltages $V_i^{\pm}$. \textbf{b}: RF voltages for the kicker (S23) and the repeller (S25) are provided separately by a different RF coil that has the same resonance frequency as the main RF coil. Using two push-pull switches the center potentials of the secondary coils can be switched synchronously and independently of each other for S23 from low to high voltages and for S25 from high to low voltages within $30\,$ns.}
\label{fig:Buncher_circ}
\end{figure*} 
 
Figure~\ref{fig:Buncher_circ} shows the electric circuit used for the operation of the buncher. A harmonic pseudopotential well is created radially between the rods of the buncher segments by applying suitable RF voltages of mirrored phases to the neighboring rods, which radially confines the ions around the center axis of the RF structure. By additionally applying a DC gradient along the segments the ions can be efficiently guided though the structure. The same working principle holds for the ion guide that is used to transfer the ions from the drift tube to the QMF. 
In the RF circuit, we utilize $100\,$nF capacitors and $100\,$k$\Omega$ resistors to couple RF and DC voltages to each of the segments, respectively. The various DC voltages are either provided directly by the CAEN multichannel power supply or indirectly via a resistor chain as depicted in Fig.~\ref{fig:Buncher_circ}(a). 
For proper operation of the buncher, we stop and store the ions in an axial potential well located at the buncher segment S24 (hereafter called the \textit{trap}) with the aim of exposing them to laser light for spectroscopy. 
After this we trigger the DC potentials of the neighbouring segments S23 (the \textit{kicker}) and S25 (the \textit{repeller}) utilizing two fast high-voltage transistor switches (type: Behlke HTS 31-06-C, $30\,$ns rise time) as depicted in Fig.~\ref{fig:Buncher_circ}(b) in order to eject the bunched ions towards the drift tube. 

\subsection{Measurement cycle and procedure} 
\begin{figure*}[htb!]
\centering
\includegraphics[width=0.92\textwidth]{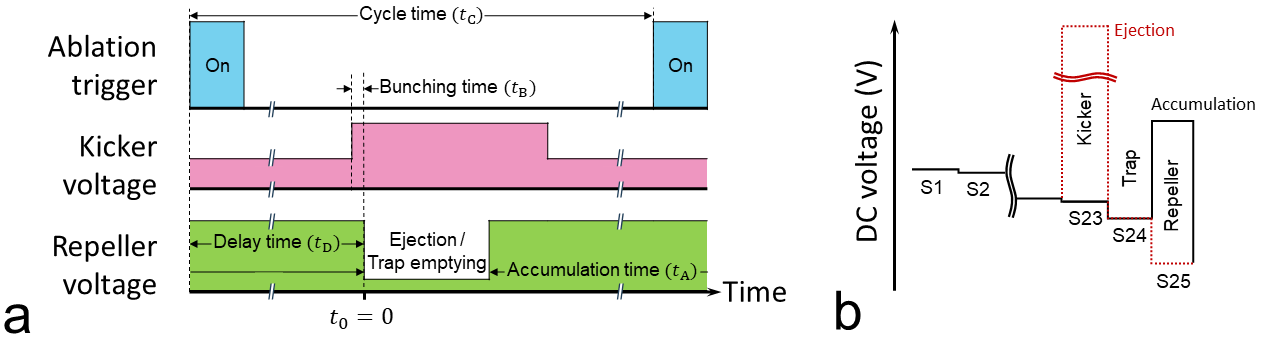}
\caption{\textbf{a}: Typical bunching cycle for electronic state chromatography. After an accumulation period, the ions are injected into the drift tube at a reference time $t_0=0$ at which the potential of the repeller segment S25 is switched from high to low values. \textbf{b}: Schematic view of the potentials applied to the different segments of the buncher. Only the potentials for S23 and S25 segments are user triggered during the cycle time. Solid (dashed) lines represent the potentials of a few buncher segments during ion accumulation (ejection). See text for more details.}
\label{fig:Cycle}
\end{figure*}

For ion bunching we first apply a retarding DC potential to the repeller (S25), which for the ablation studies was chosen to be $13\,$V higher than that of the trap (S24) for an effective accumulation of the ions at this latter trap. 
In contrast, already a $+2\,$V higher DC voltage for the repeller compared with the trap was found to be sufficient for ion trapping in the case of efficiency measurements using $^{219}$Rn$^+$ from the recoil source (\textit{cf.} Tab.~\ref{tab:Voltages1}). 
This is because ions from the stopping cell are subjected to gas collisions already during their extraction compared to those from the ablation source and are therefore thermalized before entering the buncher. 
After an appropriate delay time $t_{\text{D}}$ (ablation studies) or accumulation time $t_{\text{A}}$ (efficiency studies) as depicted in the measurement cycle in Fig.~\ref{fig:Cycle} (a), we lower the DC potential of the repeller segment for at least $50\,$µs to empty the trap and thereby let the ions be injected into the drift tube at time $t_0 = 0$, which sets the reference time for the arrival time spectra. 
$4.5\,$µs before doing so, we increase the DC potential of the kicker segment to a relatively higher value (respectively, $113\,$V and $26\,$V above the trap voltage in the ablation and efficiency studies) to further spatially confine the ions in the trap and in order to push them quickly forward to the drift tube. 
In this way we prevent ``fresh'' ions not yet stored in the trap from entering the drift tube during the ejection time. Measurements carried out utilizing such bunching cycles are referred to as \textit{bunching mode} operation.
In such a mode, we expect the ion bunch to be no wider than the adjusted $4.5\,$µs in time when it is \textcolor{black}{just} released from the trap, as the ions are continuously cooled by the surrounding background gas.
After emptying the trap the DC potential of the repeller is increased to its initial value and the accumulation cycle starts over. 
During the ablation studies, in general, the laser ablation pulse is synchronised to occur after each ion ejection phase such that newly generated ions have sufficient time ($\geq 500\,$µs) to cool down in the RF structure before bunching can take place. By operating the kicker and repeller only at their lower DC voltages, which we refer to as \textit{transmission mode}, we can transport the ions through the RF buncher without axial trapping, \textit{cf.} Fig.~\ref{fig:Cycle}(b).

\section{Commissioning of the apparatus}
\subsection{Testing of the quadrupole mass filter}
\begin{figure}
\centering
\includegraphics[width=0.47\textwidth]{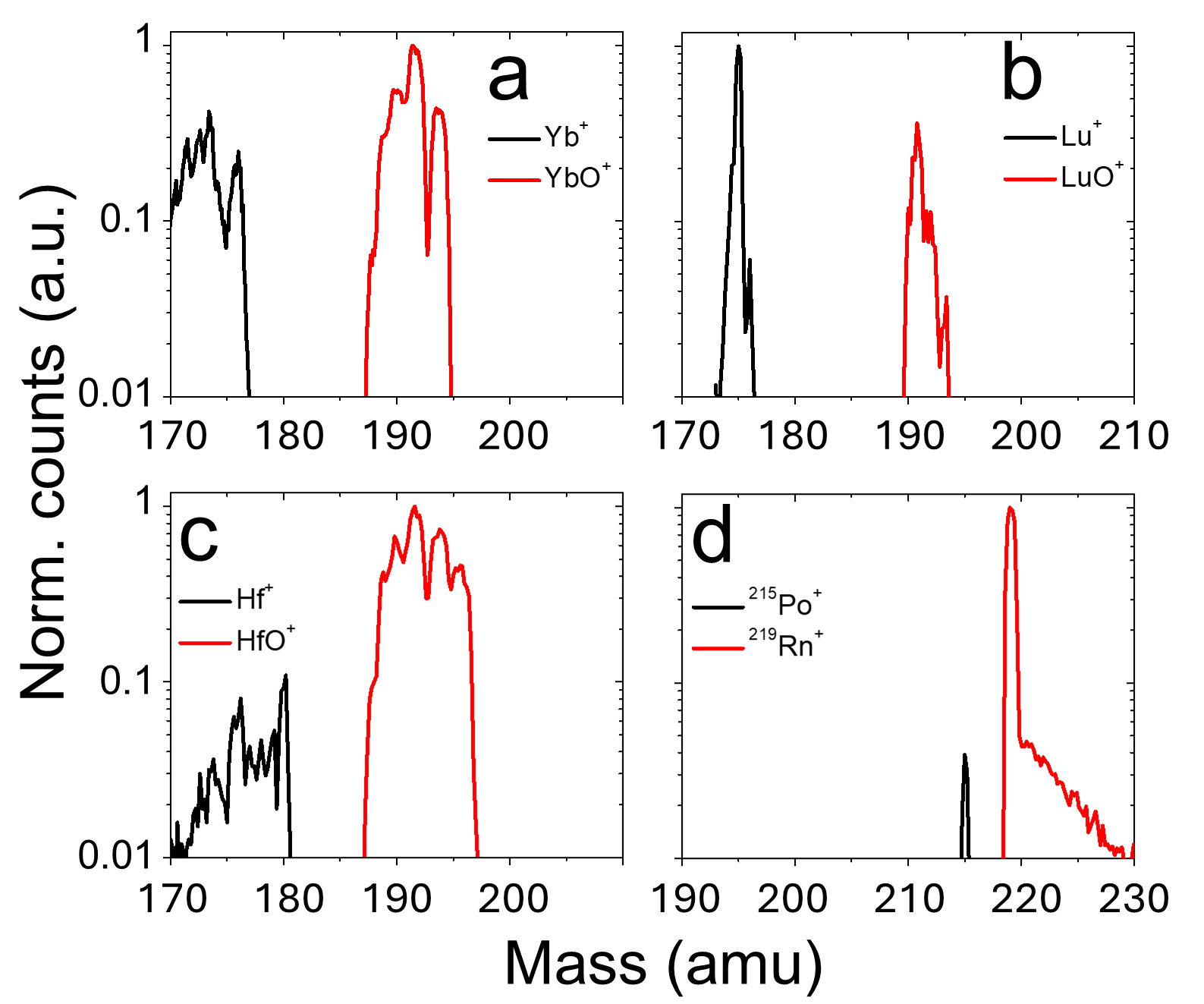}
\caption{Mass spectra obtained by laser ablation from metallic foils of \textbf{a}: Yb, \textbf{b}: Lu, \textbf{c}: Hf in vacuum, and \textbf{d}: when extracting $^{219}$Rn$^+$ and $^{215}$Po$^+$ ions from a $^{223}$Ra recoil source installed inside of the buffer-gas stopping cell. The extended tail in the mass spectrum towards higher masses in (d) is due to background alpha activity from the ions deposited on the detector.}
\label{fig:mass}
\end{figure} 
Figure \ref{fig:mass} shows mass spectra obtained for the different ion sources, including the $^{223}$Ra recoil source. The drift tube was kept evacuated during these measurements. The mass spectra obtained with laser ablation sources exhibit mass peaks for abundant isotopes of Yb$^+$, Lu$^+$, Hf$^+$, and their respective monoxides. They were recorded at a moderate mass resolving power of $m/\Delta m\approx240$. 
This QMF operation was completely sufficient for the suppression of neighbouring isotopes in the mass spectra, although the natural abundances of the elemental isotopes could not be exactly reproduced during the mass scans. 
The mass spectrum around $219$ atomic mass units (amu) was recorded with a similarly moderate resolving power. Due to the fast extraction of the recoil ions from the stopping cell~\cite{Neumayr:2006c}, not only the mass peak of $^{219}$Rn$^+$, but also that of its daughter nucleus $^{215}$Po$^+$ could still be clearly identified despite the extremely short half-life of this latter isotope of only $1.78\,$ms.

\subsection{Optimizing the bunching mode operation}
\label{sec:Bunching}
Test experiments utilizing laser ablated Hf$^+$ ions were conducted to establish and optimize the buncher operation. Initially, the drift tube was kept evacuated during these measurements to study the general trends of ion transfer through the apparatus in both transmission and bunching mode. The application of a suitable DC potential gradient along the buncher turned out to be essential for an efficient ion transport and trapping. 
At typical background pressures on the order of $10^{-2}\,$mbar, about $45\,$mV/cm DC voltage gradient proved to be most suitable for a smooth guiding of the ions to the trap. 
Once trapped in the RF structure the ions undergo collisions with the surrounding gas during the accumulation phase, lose further kinetic energy, and cool to background gas temperature at the potential well minimum. 
When gradients higher than the above-mentioned are applied, ions can escape the RF structure \textcolor{black}{due to their higher kinetic energy,} leading to ion losses. In contrast, smaller gradients cause the ions to become trapped at each of the structure segments on their way to the actual trap segment (S24).

\begin{figure}[t!]
\centering
\includegraphics[width=0.47\textwidth]{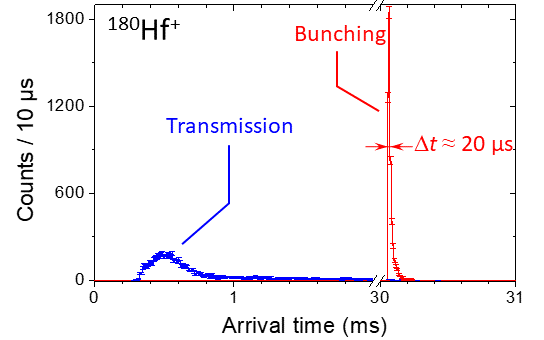}
\caption{Arrival time distributions of ablated $^{180}$Hf$^+$ in transmission (\textcolor{blue}{\textbf{---}}) and bunching (\textcolor{red}{\textbf{---}}) operation modes at a background pressure in the drift tube and in the PS2 section of $4.4\cdot 10^{-2}\,$mbar. The buncher electrode configurations are given in Tab.~\ref{tab:Voltages1}. The bunching frequency is $10\,$Hz. Unlike shown in Fig.~\ref{fig:Cycle}(a) we exceptionally take here the reference time $t_0=0$ to be the time when we trigger the ablation laser. The release of the ion bunch follows after a delay time of $t_{\text{D}}=30\,$ms. \textcolor{black}{$\Delta t$ is the peak full width at half maximum.}}
\label{fig:BunchingWorks}
\end{figure}

Figure \ref{fig:BunchingWorks} shows the two operation modes (i.e. transmission \& bunching) when applying the corresponding electrode voltage configurations as listed in Tab.~\ref{tab:Voltages1}. 
As mentioned earlier, there is no axial ion trapping in transmission mode. This way, laser ablated Hf$^+$ ions get transferred directly to the drift tube. The ions exhibiting a rather broad arrival time distribution arrive at the detector about $500\,$µs after triggering the ablation laser \textcolor{black}{(i.e., about $355\,$µs after laser ablation takes place). The peak broadening we observe in this mode of operation is caused mainly by the temporal and spatial evolution of the plasma plume during the ablation process itself.} 

When bunching is activated by triggering the DC voltages of the kicker and the repeller the ions are first trapped at segment S24 before their release towards the drift tube, in this case $30\,$ms after triggering the ablation laser. 
\textcolor{black}{The measured mean arrival time is about $64\,$µs. Although the ions are spatially confined in the trap segment before their ejection into the drift tube, the peak full-width-at-half-maximum ($\Delta t$) amounts to $\approx20\,$µs. We attribute this peak broadening to residual longitudinal diffusion during ion drift inside of the drift tube and Rf ion guide at a background pressure of $4.4\cdot 10^{-2}\,$mbar.}

The optimum trapping time is a compromise between ion accumulation and loss in the RF fields, ion confinement in space for precise bunching in the time domain, and possible quenching of the ionic metastable state of interest due to gas collisions. 

The number of ions that can be trapped depends on the buncher geometry and on the axial depth of the potential well that exists between the kicker and the repeller segments. In our experiments we measure the highest ion transmission in the bunching mode when the axial potential of the trap is set equal to that of the kicker. 
Such a constellation results in an axially elongated potential well with a relatively large storage volume, which would, however, lead to broader arrival time spectra.
To achieve optimal chromatography, we therefore apply a $1$-V lower DC potential to the trap than to the kicker to confine the ions in the trap where they shall be exposed to resonant laser excitation in LRC experiments.

Furthermore we found out that buffer-gas cooling of the ions inside of the buncher 
at background pressures on the order of $10^{-2}\,$mbar
is crucial to substantially reduce ion losses. 
In contrast to the efficiency measurements (\textit{cf.} Sec. \ref{sec:Efficiency}), the ion transport in the ablation studies in addition suffered from inefficient catching of the ions in the RF fringing fields of the buncher, which one can partially counteract by increasing the PS2 background pressure beyond $8\cdot10^{-2}\,$mbar. In Sec.~\ref{sec:Efficiency} we give more details on the efficiency aspects of the LRC apparatus. 
\begin{figure}[b!]
\centering
\includegraphics[width=0.47\textwidth]{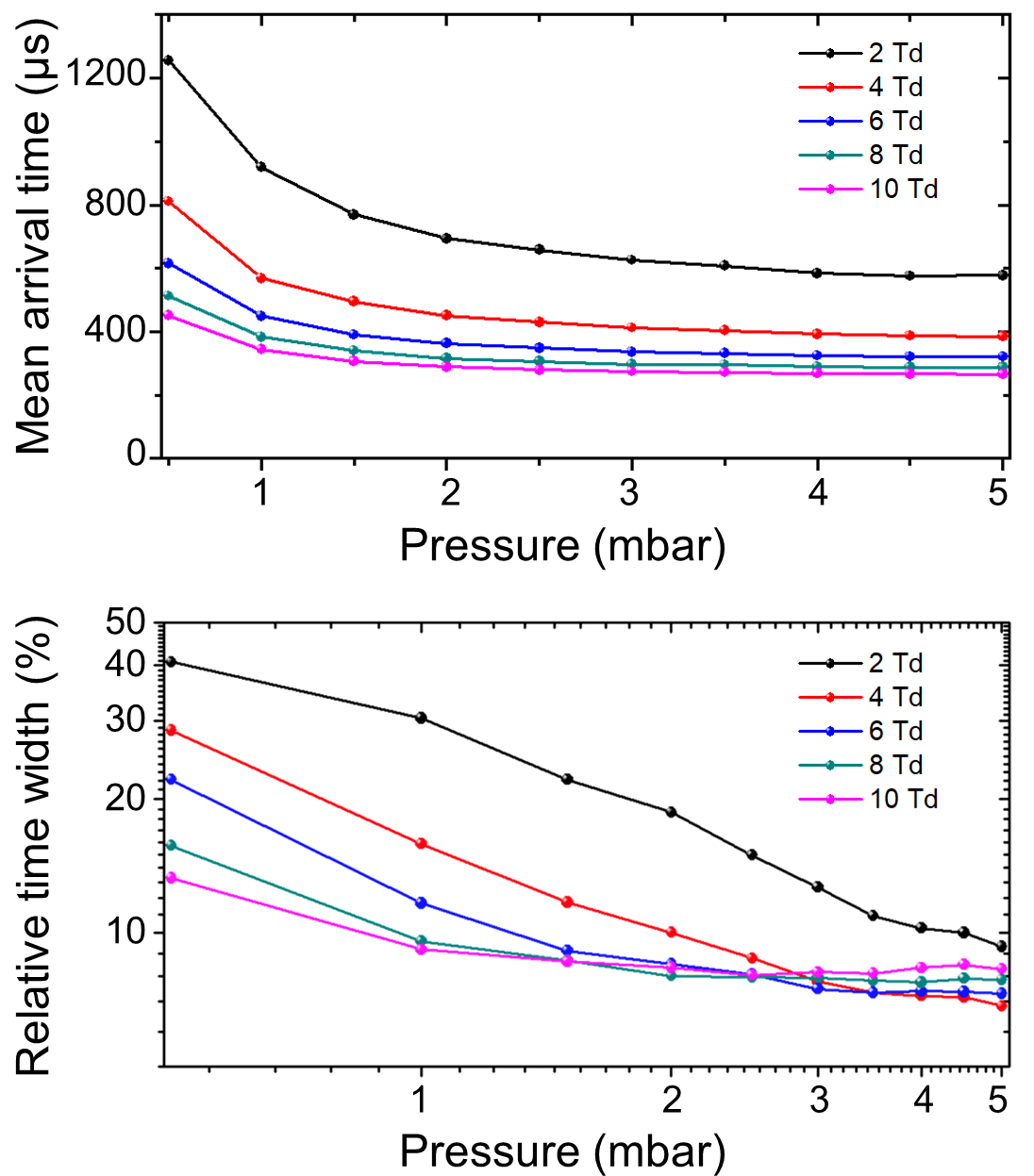}
\caption{Mean arrival time \textit{\~t} of ablated Hf$^+$ ions (upper panel) and the relative time width $\Delta t/$\textit{\~t} (lower panel) as function of the drift tube pressure for different reduced electric fields from $2$ to $10\,$Td. See text for more details.}
\label{fig:Pressure_sigma}
\end{figure} 

In order to assess the chromatography performance of the apparatus at room temperature, we first studied $^{180}$Hf$^+$ arrival time at different drift tube pressures from $0.5$ to $5\,$mbar and reduced electric fields $E/n_0$ from $2$ to $10\,$Td, with $1\,$Td (Townsend unit) being equal to $10^{-21}\,$Vm$^2$. The corresponding mean arrival time (\textit{\~t}) is shown in Fig.~\ref{fig:Pressure_sigma}. 
This is composed of the drift time $t_{\text{d}}$, which according to Eq. \ref{Eq:velocity} is inversely proportional to the reduced fields, and an offset time of flight ($t_{\text{ToF}}$) the ions need to traverse the distance from the drift tube to the channetron detector.
A steady-state ion drift can be assumed when the terminal velocity of the ion is reached, i.e. when the drift time does not significantly change at a fixed reduced field $E/n_0$.
Since the cryogenic drift tube in use is extremely short this velocity is reached only at low-field conditions and at drift pressures $\gtrsim 3\,$mbar.
This fact can be inferred from the stagnation of the mean arrival times at higher pressure values as shown in the upper panel of Fig. \ref{fig:Pressure_sigma}. In the lower panel of Fig. \ref{fig:Pressure_sigma} we show the corresponding relative width $\Delta t/$\textit{\~t} of the ATDs, which is defined as the ratio of the full-width-at-half-maximum of the time peak ($\Delta t$) to the mean arrival time \textit{\~t}. 
It decreases with increasing pressure until it is below $10$\% above $3\,$mbar and reaches a minimum of about $7$\%, corresponding to a resolving power of \textit{\~t}$/\Delta t\approx14$ at a reduced electric field of $4\,$Td.
In fact, applying higher drift pressures should enable even higher resolving power of the drift tube. In our experiments, however, we applied only of up to $5\,$mbar drift pressures, to enable a proper operation of the mass filter due to limited PS2 pumping capacities. 
In the following, and in order to achieve the required temporal resolution for the separation of electronic states of different configurations without risking metastable state deactivation at higher pressures, we generally operated the drift tube between $3$ and $4\,$mbar. \textcolor{black}{We thereby keep the time resolving power a free parameter, which we adapt to the requirements of each case under study.}

\subsection{Chromatography performance}
Electronic state chromatography is a well-studied effect and has been reported for many transition metal ions drifting in helium~\cite{Viehland:2018}. The effect becomes most pronounced and relatively easy to observe when completely different electronic configurations of the ions are present \cite{Armentrout:2011}. Thus, for the start-up phase we investigated Yb$^+$ and Lu$^+$ ions to assess the chromatography performance of the apparatus.
Yb$^+$ possesses the electronic configuration $4f^{14} 6s^1\,^2$S$_{1/2}$ in its ground state, whereas Lu$^+$ exhibits a doubly occupied $6s$ orbital with a ground-state configuration $4f^{14} 6s^2\,^1$S$_0$. 
Since the double occupation of the $s$ orbital would lead to a more intense interaction of the ion with helium atoms we expect Lu$^+$ to be relatively slow in its ground state with respect to Yb$^+$ as has been predicted by theory \cite{Visentin:2020} and observed in former measurements at a He gas temperature of about $295$K~\cite{Manard:2017}. The ion mobility difference was found to be $14$\%, which is sufficiently large enough to be resolved utilizing our cryogenic drift tube.
In addition, at gas temperatures above $100$K the mobility of Yb$^+$ is expected to be similar to that of Lu$^+$ in its lowest metastable state with the configuration $4f^{14} 5d^1 6s^1$ due to the singly occupied $s$ orbital in both cases \cite{Visentin:2020, Laatiaoui:2020b}. 
\begin{figure}[tb!]
\centering
\includegraphics[width=0.5\textwidth]{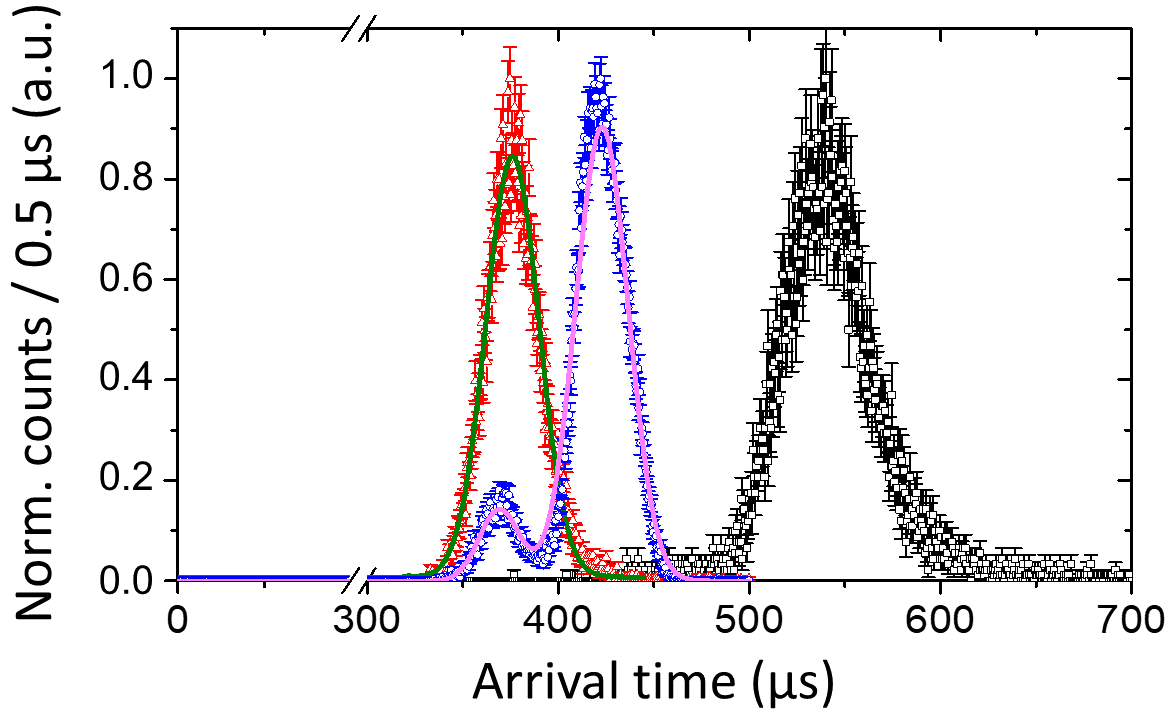}
\caption{Superimposed and normalized arrival time distributions for $^{176}$Yb$^+$ (\textcolor{red}{$\vartriangle$}), $^{175}$Lu$^+$ (\textcolor{blue}{$\circ$}), and $^{175}$LuO$^+$ (\textcolor{black}{$_\square$}) ions after their drift in helium at a pressure of $4\,$mbar under the influence of a reduced electric field of $4\,$Td. Solid lines indicate the corresponding Gaussian fits to the data. The measurement time amounts to $t_{\text{M}}= 5\,$min and the gas temperature $T \approx 295\,$K. Other parameters are $t_{\text{C}}= 100\,$ms and $t_{\text{B}} = 4.5\,$µs.}
\label{fig:Chromatography}
\end{figure} 
\begin{figure}[tbh!]
\centering
\includegraphics[width=0.47\textwidth]{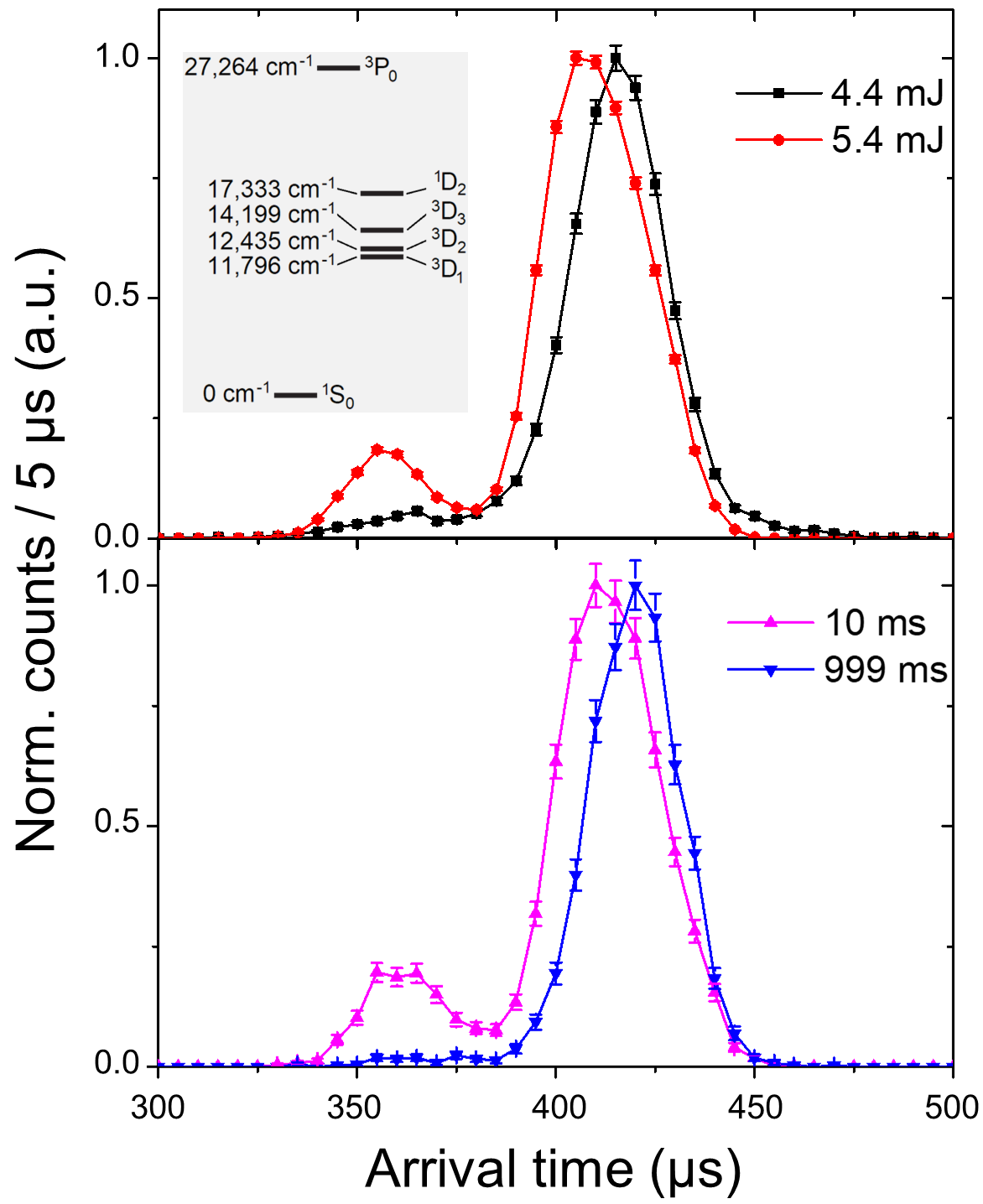}
\caption{Superimposed and normalized arrival time distributions for $^{175}$Lu$^+$ ions drifting in helium at $E/n_0=4\,$Td. 
P$ = 4\,$mbar, $t_{\text{M}}= 5\,$min, $t_{\text{B}} = 4.5\,$µs.
\textbf{a}: Laser pulse energy for ablation was decreased from $5.4\,$mJ (\textcolor{red}{$\bullet$}) to $4.4\,$mJ ($_\blacksquare$) at $10\,$Hz ablation frequency and $t_{\text{C}}= 100\,$ms cycle time. $t_{\text{D}} = 500\,$µs. Inset: Part of the level scheme of Lu$^+$ with the relevant energy levels for the ground state and the lowest metastable states.
\textbf{b}: Delay time between ablation and ion ejection trigger was increased from $t_{\text{D}} = 10\,$ms (\textcolor{Rhodamine}{$\blacktriangle$}) to $t_{\text{D}} = 999\,$ms (\textcolor{blue}{$\blacktriangledown$}) at $5.4\,$mJ laser pulse energy and $1\,$Hz ablation frequency. $t_{\text{C}}= 1\,$s. See text for more information.}
\label{fig:Cooling_time}
\end{figure} 

Figure~\ref{fig:Chromatography} shows measured and superimposed arrival time distributions of $^{176}$Yb$^+$ and $^{175}$Lu$^+$ that were produced by the ablation source at an ablation frequency of $10\,$Hz and a relatively high laser pulse energy of $5.4\,$mJ. The measurements were taken successively at the same boundary conditions of $4$\,Td reduced electric field and $4.5\,$µs bunching time. For comparison we also show the ATD of LuO$^+$ molecular ions superimposed to the other ATDs. The bulky monoxide molecules exhibit larger collision cross sections when interacting with He and thus need about $120\,$µs more time to traverse the drift tube compared with Lu$^+$ in the ground state. 
Remarkably however is the appearance of a second peak centered around $369\,$µs in the ATD spectrum of Lu$^+$, which should be related to the arrival time of Lu$^+$ ions in the lowest metastable state. Here, one should note that Manard and co-workers did not observe such a ``faster'' Lu$^+$ ion in high-precision ion mobility experiments \cite{Manard:2017}. They traced back the absence of the metastable peak to either a breakdown of the Russell-Saunders coupling or --what we think to be the major reason-- an efficient quenching of electronic excited states during ion drift, using a relatively long ($2\,$m) drift tube back then.

In our experiments and for a firm assignment of the smaller peak to excited lutetium ions, we reduced the population of the metastable state by reducing the pulse energy for laser ablation. 
Figure~\ref{fig:Cooling_time}(a) shows such a scenario when we reduced the laser pulse energy from $5.4\,$mJ to $4.4\,$mJ. For a better qualitative comparison, we normalized the heights of the ATD peaks to $1$, since ion production is very inefficient at low ablation power. 
As can be seen in Fig.~\ref{fig:Cooling_time}(a) and as expected, such a reduction in pulse energy results in relatively fewer counts being registered at the position of the suspected metastable peak.
Both the ground-state peak shape and center position are found to depend on the pulse energy. We trace back this effect to population of higher-lying metastable states such as the $^1$D$_2$ fine-structure component at elevated pulse energies, \textit{cf.} inset in Fig.~\ref{fig:Cooling_time}(a). A rapid deactivation of this state to the ground state in collisions with He inside of the drift tube would lead to an abrupt change in the velocity of the ions during their drift and thus to a slight change in the peak shape. 
Consequently and in order to ensure optimum time resolution, optical resonance pumping should be performed in future LRC experiments only when the steady state of ion production and chromatography has been reached.
In a second test experiment we kept the pulse energy constant at $5.4\,$mJ and reduced the ablation frequency to $1\,$Hz such that different delay times can be applied. 
Figure \ref{fig:Cooling_time}(b) shows superimposed ATDs of Lu$^+$ when the ions are accumulated and cooled in the RF trap for different delay times. Depending on this time, the fraction of excited ions within the trapped ion ensemble changes due to collisional quenching. If the delay time is kept short with $t_{\text{D}} = 10\,$ms, the ions are quickly injected into the drift tube after ablation and we measure a substantial metastable fraction of about $14$\%.
We also show in Fig.~\ref{fig:Cooling_time}(b) a scenario in which the ions are trapped for an extremely long time ($t_{\text{D}} = 999\,$ms). Still about $1$\% of the ions ``survive'' gas collisions in their metastable state even at a background pressure of about $3 \cdot 10^{-2}\,$mbar.
From these measurements we deduce a quenching rate of the Lu$^+$ metastable state to be \textcolor{black}{$\alpha = (8.8\pm0.2) \cdot 10^{-14}\,n_0\,$cm$^3$/s} at room temperature and low reduced fields. 

Although the D multiplet components might have similar time fingerprints in the ATDs due to the same configuration they share, we expect collision-induced intramultiplet transitions to efficiently populate the lowest-lying $^3$D$_1$ state due to the small energy gaps between the different levels, \textit{cf.} inset in Fig.~\ref{fig:Cooling_time}(a). For this reason, we tentatively assign the metastable peak to this extremely long-lived ($^3$D$_1$) state with a predicted lifetime of $54\,$h \cite{Paez:2016}.

Furthermore, from Fig.~\ref{fig:Chromatography} one may suspect similar mobilities for Yb$^+$ in the ground state and Lu$^+$ in the metastable state because both ionic species are detected at similar mean arrival times under same boundary conditions of gas temperature and reduced fields. As the drift pressure of $5\,$mbar could not be exceeded in our experiments and the drift tube is extremely short and not optimally suited for accurate mobility measurements, we only report here on relative mobility differences for the investigated species. These differences can be expressed as~\cite{Laatiaoui:2020b}
\begin{equation}\label{deltaTtd}
(K'-K)/K' = (t_{\text{d}}-t'_{\text{d}})/t_{\text{d}}
\end{equation}
with $K'$ and $t'$ denoting the mobility and the drift time of a faster reference species, respectively. 

Since the drift time is extracted from the mean arrival time according to $t_{\text{d}} = \text{\textit{\~t}} - t_{\text{ToF}}$, we first determined the ion time of flight for each drift pressure from linear extrapolations of measured \textit{\~t} towards higher electric fields. In addition to the $t_{\text{d}}$ obtained in this way, it turned out that the time of flight depends on the species investigated, albeit only slightly. We attribute this latter dependence to the residual drift within the RF ion guide at elevated background pressures. For Lu$^+$ in the ground state, for example, it is $(190\pm3)\,$µs, while for Yb$^+$ it is only $(180\pm2)\,$µs. Finally, according to equation~\ref{deltaTtd}, we obtain a relative mobility difference of $14(2)$\% between Lu$^+$ and Yb$^+$ at $T\approx295$K, in excellent agreement with previous experimental findings \cite{Manard:2017}. In the case of Lu$^+$ in its different electronic states this relative difference amounts to $19(4)$\%, which is still consistent with different theory predictions~\cite{Laatiaoui:2020b,Ramanantoanina:2023}.
  
\subsection{Efficiency aspects}\label{sec:Efficiency}
\begin{figure}[b!]
\centering
\includegraphics[width=0.47\textwidth]{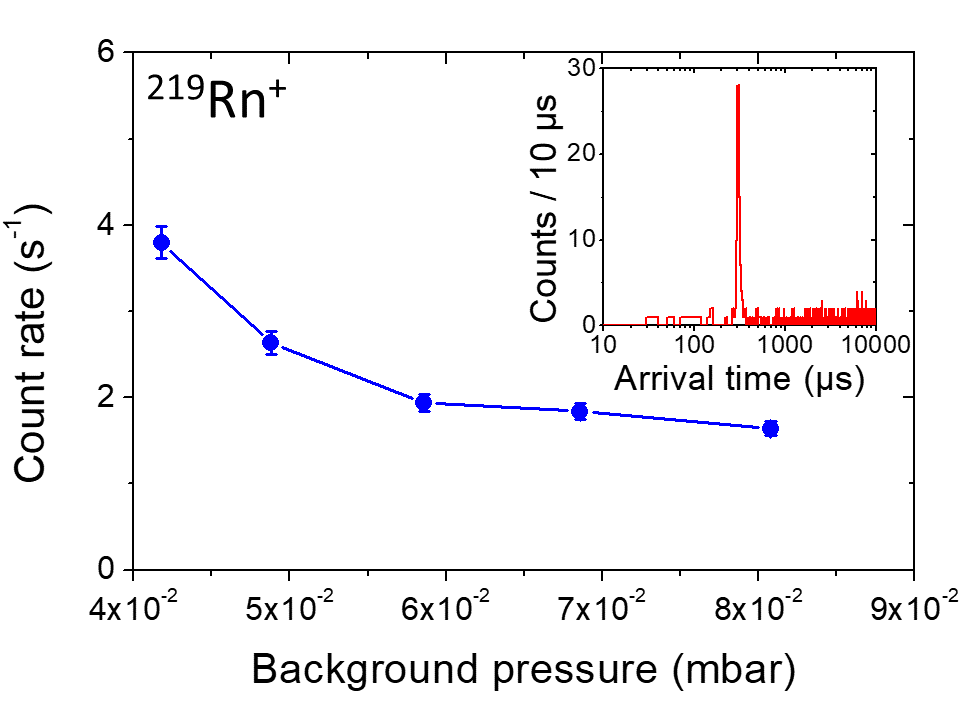}
\caption{Optimization of the transmission and bunching operation modes utilizing $^{219}$Rn$^+$ ions. Shown are Rn$^+$ count rates measured at different background pressures in the PS2 vacuum section. Both the stopping cell and the drift tube pressures were kept constant at about $63\,$mbar and $3.1\,$mbar, respectively. Inset: Bunching of $^{219}$Rn$^+$ ions at a background pressure of $5\cdot 10^{-2}\,$mbar, a bunching frequency of $100\,$Hz, a drift tube pressure of $P\approx 3.4\,$mbar, and a gas temperature of $T=295$K. The background counts are due to residual activity from implanted $^{219}$Rn and its daughters.}
\label{fig:Efficiency}
\end{figure} 
For the efficiency studies, we first optimized the ion transport utilizing ablated lutetium ions as mentioned before. Then we used a $^{223}$Ra recoil source to further optimize and subsequently quantify the overall efficiency of the apparatus in both transmission and bunching mode.
To this end we installed the source inside of the buffer-gas stopping cell as shown in Fig.~\ref{fig:LRC_setup}. 
Its alpha activity $A(^{223}\text{Ra})$ has been determined beforehand by alpha spectroscopy under vacuum conditions. 
The voltage configuration used during the efficiency studies is shown in Tab.~\ref{tab:Voltages1}. From the preparation procedures of such recoil sources, it can be safely assumed that the radium atoms remain attached to the surface of the source and thus that $50$\% of the $^{219}$Rn daughter ions can easily escape into the buffer gas and be extracted. 

Figure \ref{fig:Efficiency} shows registered Rn$^+$ count rates in transmission mode at a stopping cell pressure of $63\,$mbar and a drift tube pressure of $3.1\,$mbar as function of the background pressure in the PS2 vacuum section. This latter was adjusted by accommodating the rated speed of the corresponding turbo-molecular pump. The increasing of the count rates at lower background pressures points to an enhancement in the gas jet formation in front of the nozzle, which helps in better catching ions into the fringing RF fields of the buncher. Due to high gas load and limited pumping capacities a background pressure $<4\cdot 10^{-2}\,$mbar could not be reached in this latter section under typical operation conditions. However, we expect the transmission efficiency to quickly saturate at lower background pressures before it decreases since an efficient radial ion cooling due to gas collisions is mandatory for best ion transport through the different RF structures. 

The overall efficiency of the setup, defined as the fraction of all source ejected ions that successfully hit the detector, can be estimated as follows: From the knowledge of the registered count rate $R$ when the mass filter is set to transmit only $^{219}$Rn$^+$ ions, the time $t_{\text{t}}$ needed to transport the ions from the stopping cell to the detector, and the time $\delta t$ elapsed since the determination of $A(^{223}\text{Ra})$ we can give an overall efficiency $\varepsilon_{\text{tot}}$ of the apparatus in the bunching mode as 
\begin{equation}
    \varepsilon_{\text{tot}}=\frac{[R \cdot \varepsilon_{\text{d}}^{-1} \cdot e^{t_{\text{t}}\ln2/T_{1/2}(^{219}\text{Rn})} \cdot f^{-1}_{\text{rad}}] - \textcolor{black}{\gamma} B} {A(^{223}\text{Ra}) \cdot \varepsilon_{\Omega} \cdot e^{-\delta t\ln2/T_{1/2}(^{223}\text{Ra})}}.
    \label{Eq_eff}
\end{equation}
Here, we assume an efficiency for detecting ions by the channeltron to be $\varepsilon_{\text{d}}=100$\% and further that the measurement time in the bunching mode is negligibly small compared to the $^{223}$Ra half-life of $T_{1/2}(^{223}\text{Ra})=11.44\,$d. As mentioned before, the efficiency for ejection and implantation of $^{219}$Rn recoil ions into the buffer gas is assumed to be $\varepsilon_{\Omega}=50$\%. We use the correction factor $f_{\text{rad}}$ to account for registered background counts within the measurement time $t_{\text{M}}$ due to radioactive decay of $^{219}$Rn and of its daughters on the channeltron detector. This amounts to \cite{Schwamb:1996}
\begin{align}
    f_{\text{rad}} = 1 + \textcolor{black}{\gamma} & \{0.535 \cdot (1-e^{-t_{\text{M}}\ln2/T_{1/2}(^{219}\text{Rn})})\nonumber\\
                & + 0.324 \cdot (1-e^{-t_{\text{M}}\ln2/T_{1/2}(^{211}\text{Pb})})\}
    \label{Eq_eff_fac}
\end{align}
\noindent
with $T_{1/2}(^{211}\text{Pb})=0.6\,$h. 
Known residual activity on the detector due to preceding measurements is taken into account by adjusting an offset parameter $B$ in Eq.~\ref{Eq_eff} to the experimental conditions for each measurement. 
Both, the correction factor $f_{\text{rad}}$ and the offset rate $B$ are relevant mainly for calculating the overall efficiency in the transmission mode. In this case, \textcolor{black}{the coefficient $\gamma$} in Eqs.~\ref{Eq_eff} \& \ref{Eq_eff_fac} is set equal to $1$. In the bunching mode operation, in particular at low bunching frequencies, they can be safely neglected because the random decay events registered by the detector are homogeneously distributed in the time domain within a typical measurement cycle of $t_{\text{C}}=10\,$ms and may contribute with a fraction of only \textcolor{black}{$\gamma \approx 5\cdot10^{-3}$} to rates associated with the $^{219}$Rn$^+$ ion bunches. The inset of Fig.~\ref{fig:Efficiency} shows an example arrival time distribution of $^{219}$Rn ions recorded at a bunching frequency of $100\,$Hz. 

At a typical helium pressure of $3\,$mbar in the drift tube and a moderate mass resolving power of the QMF, the overall efficiency of the apparatus in the transmission mode is found to be $(2.8\pm0.2)\%$. In the bunching mode operation this amounts to $(0.6\pm0.1)\%$ and is found to be rather independent of the bunching frequency up to $1\,$kHz. 
The main efficiency losses are caused by the ion extraction itself and, above all, by diffusion losses in the drift tube. The efficiency of ion extraction is $30$\%~\cite{Droese:2014}. For the drift tube, the transmission efficiency has not yet been measured, but earlier SIMION simulations for Sc$^+$ ions drifting in He indicated significantly lower efficiencies~\cite{RomeroRomero:2022}, ranging from $5$\% to $20$\%, depending on the applied reduced fields. In general, the higher the field the higher the ion transmission. However, higher electric fields are not always desirable because metastable states tend to diabatically quench to the ground state in energetic collisions due to level crossings. Thus, the trade-off we made in operating the drift tube at low reduced fields may explain the transmission losses during the inauguration experiments. 

\section{Summary and Outlook}
A laser resonance chromatography apparatus has been developed and put into operation for first offline experiments on Lu$^+$. In the start-up phase the corresponding cryogenic drift tube was operated with helium gas at a temperature of about $295$K and reduced fields as high as $10\,$Td. The chromatography performance was assessed by analyzing ATDs of laser ablated Hf$^+$ ions and ATD peak separations when comparing Lu$^+$ with Yb$^+$ ions in their ground states. A metastable ATD peak was observed for the first time in Lu$^+$ arrival time distributions. The measured relative mobility difference of this ion in its ground and metastable state is about $19$\%. The relative difference between Lu$^+$ and Yb$^+$ in their ground states is $14$\%, which is in excellent agreement with former findings~\cite{Manard:2017}. In addition, we determined the rate for deactivation of the $^3$D$_1$ metastable state in Lu$^+$ drifting at low fields in He to be \textcolor{black}{$(8.8\pm0.2) \cdot 10^{-14}\,n_0\,$cm$^3$/s}. Moreover, utilizing $^{219}$Rn recoil ions provided from a sample $^{223}$Ra source, we measured an overall efficiency of the apparatus of $0.6$\% in the bunching mode, which is the operation mode to be applied in future LRC experiments.

The next steps include investigating the resonant laser excitation of Lu$^+$ ions and how this affects the chromatography signal. The focus is on the $^1$S$_0$--$^3$P$_1$ optical transition in this ion at about $28,503\,$cm$^{-1}$, which enables optical pumping to the $^3$D$_1$ metastable state \cite{Laatiaoui:2020a}. Systematic studies have to be carried out to elucidate the influence of laser intensities on the broadening of spectral lines and the achievable accuracy in determining states transition strengths, lifetimes, and hyperfine constants. Furthermore and in order to demonstrate the applicability of the method for optical spectroscopy of transition metal cations, various commercially available chemical elements were selected for initial offline runs, such as scandium, vanadium, and tungsten, for which electronic state chromatography has already been demonstrated, to name but a few.
Prior to performing future studies at in-flight separator facilities it will be necessary to further optimize the LRC technique in terms of exploring the spectral precision of the method and enhancing the setup's overall efficiency in the bunching mode. One of the obvious and chromatography-independent improvement will be the use of a cryogenic stopping cell, which enables an increase in ion extraction efficiency by approximately a factor of $2.5$ compared to the currently used buffer gas stopping cell~\cite{Droese:2014,Kaleja:2019}.

\section{Acknowledgements}
We thank J. Diefenbach, D. Rennisch, and C. D\"ullmann for providing the $^{223}$Ra recoil source for calibration purposes as well as W. Lauth for designing the electric circuit of the buncher. \textcolor{black}{B.J. acknowledges funding from the Deutsche Forschungs Gemeinschaft (DFG, German Research Foundation) (Project No. 426500921).}
This project has received funding from the European Research Council (ERC) under the European Union’s Horizon 2020 Research and Innovation Programme (Grant Agreement No. 819957).

\bibliography{main}
\end{document}